# Cyber Security of Smart Grid Infrastructure

Adnan Anwar, Abdun Naser Mahmood
School of Engineering and Information Technology
University of New South Wales
Canberra, Australia

**Abstract:**

Smart grid security is crucial to maintain stable and reliable power system operation during the contingency situation due to the failure of any critical power system component. Ensuring a 'secured smart grid' involves with a less possibility of power grid collapse or equipment malfunction. Due to lack of the proper 'security measures', a major blackout may occur which can even lead to a cascading failure. Therefore, to protect this critical power system infrastructure and to ensure a reliable and an uninterrupted power supply to the end users, smart grid security issues must be addressed with high priority.

In a smart grid environment, electric power infrastructure is modernized by incorporating the current and future requirements and advanced functionalities to its consumers. To make the smart grid happen, cyber system is integrated with the physical power system. Although adoption of cyber system has made the grid more energy efficient and modernized, it has introduced cyber-attack issues which are critical for national infrastructure security and customer satisfaction. Due to the cyber-attack, power grid may face operational failures and loss of synchronization. This operational failure may damage critical power system components which may interrupt the power supply and make the system unstable resulting high financial penalties. In this chapter, some recent cyber attack related incidents into a smart grid environment are discussed. The requirements and the state of the art of cyber security issues of a critical power system infrastructure are illustrated elaborately.

## 1 Introduction: Cyber Attacks on Smart Grid

In recent years, power system has faced several cyber related attacks which have raised the question regarding the security vulnerabilities and its large scale impact on the critical power system infrastructure. Some significant issues related to cyber-attack on the power grid are discussed in the following section.

1. In the middle of 2010, a computer worm 'Stuxnet' was discovered which spreads using 'Windows' operating system and targets Siemens industrial software and equipment to unstable power system operation [1]. This type of cyber-attack based on the intrusion of computer virus targeting industrial power plant introduces new threads to both cyber and physical systems [2].

2. On August 14, 2003, large portions of the Midwest and Northeast United States and Ontario, Canada, experienced an electric power blackout which remained for up to 4 days in some parts by affecting around 50 million people and 61,800 megawatts (MW) of electric load in some parts of the United States [3]. Although this historical large scale blackout is not directly related to malicious activity of the cyber terrorists, it is caused by a failure in the software program of the cyber system [3].

3. On September 28, 2003, Italy and some parts of Switzerland faced its largest power supply disruption affecting 56 million people in total [3]. This blackout is restored after 18 hours in Italy resulting huge financial loss. The blackout happened because of the technical difficulties caused by the human error and Ineffective communication within the power grid operators.

4. Another large blackout occurred in the south west Europe due to the human error on November 04, 2006 [3]. Insufficient communication was also an important issue behind the blackout.

5. According to the 2011 annual report of the Repository for Industrial Security Incidents (RISI), around 35% of industrial control system (ICS) security incidents were instigated through the remote access within the cyber system [4,5]. Power and utility sector faced around 12 cyber security incidents between 2004 to 2008 which is around 20% increase of this type of cyber incidents compared with the previous 4 years [4]. As ICS and SCADA is playing a vital role in a smart grid infrastructure, the cyber security concern in increasing rapidly.

From the above discussion, it can be seen that some major cyber-physical vulnerabilities of the smart grid are related to the cyber issues. Therefore, Smart Grid Infrastructure Security (SGIS) must address the deliberate attacks by the cyber terrorists and industrial espionage, disgruntled employees, user errors, equipment failures, and natural disasters [6]. In order to protect the critical smart grid infrastructure, anomaly detection can play a vital role by identifying malicious data in the network.

The recent incidents related to cyber-attacks in a smart grid were discussed in this section which motivates the power system, communication and computer engineers and researchers for future research in this challenging area. In Section 2, definitions and characteristics of a cyber-physical smart grid are discussed. Different aspects affecting the smart grid security are presented in Section 3. In Section 4, security requirements of an attack resilient smart grid are discussed. Some possible smart grid anomalies are described and reviewed in Section 5. In Section 6, protection techniques of smart grid from cyber-attacks are discussed. Finally, Section 7 concludes with references.

## 2 Smart Grid Infrastructures

### 2.1 Definition of a Smart Grid

The Smart Grid concept is evolved to make the power grid more energy efficient and intelligent. According to the US Department of Energy, smart grid can be defined as:

*"Smart grid generally refers to a class of technology people are using to bring utility electricity delivery systems into the 21st century, using computer-based remote control and automation. These systems are made possible by two-way communication technology and computer processing that has been used for decades in other industries. They are beginning to be used on electricity networks, from the power plants and wind farms all the way to the consumers of electricity in homes and businesses. They offer many benefits to utilities and consumers -- mostly seen in big improvements in energy efficiency on the electricity grid and in the energy users' homes and offices."*

Traditionally, power grid was designed to transport power from the generation plant to the end-users. Therefore, the whole power flow pattern was uni-directional and the control structure was centralized. In order to take advantage of the advanced technology to control power flow and to mitigate the ever-growing load demand, new communication techniques and distributed energy resources are being incorporated within the physical power system infrastructure. Integration of distributed generations has introduced bi-directional power flows into the grid . Moreover, energy storage devices, plug-in hybrid electric vehicles, and other advanced physical components of the power system have introduced more complexity into the grid. On the other hand, deployment of the communication network (e.g., SCADA system and Advanced Metering Infrastructure) has provided more stability, reliability, flexibility and efficiency in the operation and control of this complex power system. However, increasingly the vulnerabilities of the physical system are being exposed due to malicious attacks on the cyber-physical smart grid infrastructure. Therefore, there is a need to identify the smart grid security issues related to cyber security.

### 2.2 Ideal Functionalities of a Smart Grid

Smart grid is the modernization of the traditional power grid which should ideally have some advanced functionalities [7]:

- Self-healing
- Motivates and includes the consumer
- Resists attack
- Increases power quality
- Accommodates all generation and storage options
- Enables electrical markets
- Optimizes assets and operates efficiently

Generally, power system is a massive and complex system and very vulnerable in terms of physical or cyber-attack. The term 'self-healing' signifies the ability of the modern smart grid to recover the situation and become stable after facing any interruption. According to the European technology platform of the Smart Grid, 'Self-healing' network not only addresses automated network restoration strategies considering distributed energy resources, but also deals with high level decentralized control methodologies to prevent blackouts [8]. Another key characteristic of smart grid is attack resiliency due to the cyber-physical attack of the grid. Power system can be treated as one of the major key public infrastructures. Therefore, damage of any component of the grid may cause enormous loss in terms of country's economy and social welfare. It is very important to protect the power grid infrastructure. In a smart grid environment, it is expected that the cyber-physical system would be attack resilient which will help to protect the country's asset and ensure national security.

## 3 Security issues of Cyber-Physical smart grid

In a smart grid, the physical power system and the cyber system of information and communication technologies are highly coupled which introduced new security concerns [9]. Smart grid security issues need to address new challenges for a reliable, safe, efficient and stable operation of the grid. It is important to note that current security approaches are either inapplicable, not viable, insufficiently scalable, incompatible, or simply inadequate which need to be replaced by new and advanced techniques to ensure the security of the highly massive and complex dynamic smart grid environment [9].

A smart grid can be treated as the combination of physical power system components and cyber system infrastructure including software, hardware and communication requirements [10]. In Figure 1, a typical smart grid architecture is shown where power will flow from bulk generation plant to end users. On the other hand, information flow will occur in both directions, i.e., in the device level for coordination and among the operators and service provider level for efficient and advanced control. Therefore, in a smart grid, both cyber and physical system securities are crucial and consideration of security issues in cyber domain and physical power system in isolation cannot capture the whole picture. According to [10], the security issues of a cyber-physical smart grid comprise of the following issues:

1. The physical components of the smart grid
2. Control centres and control applications
3. The cyber infrastructures for smart grid stable, reliable, and efficient operation and planning
4. The correlation between cyber-attacks and the resulting physical system impacts
5. The protection measures to mitigate risks from cyber threats

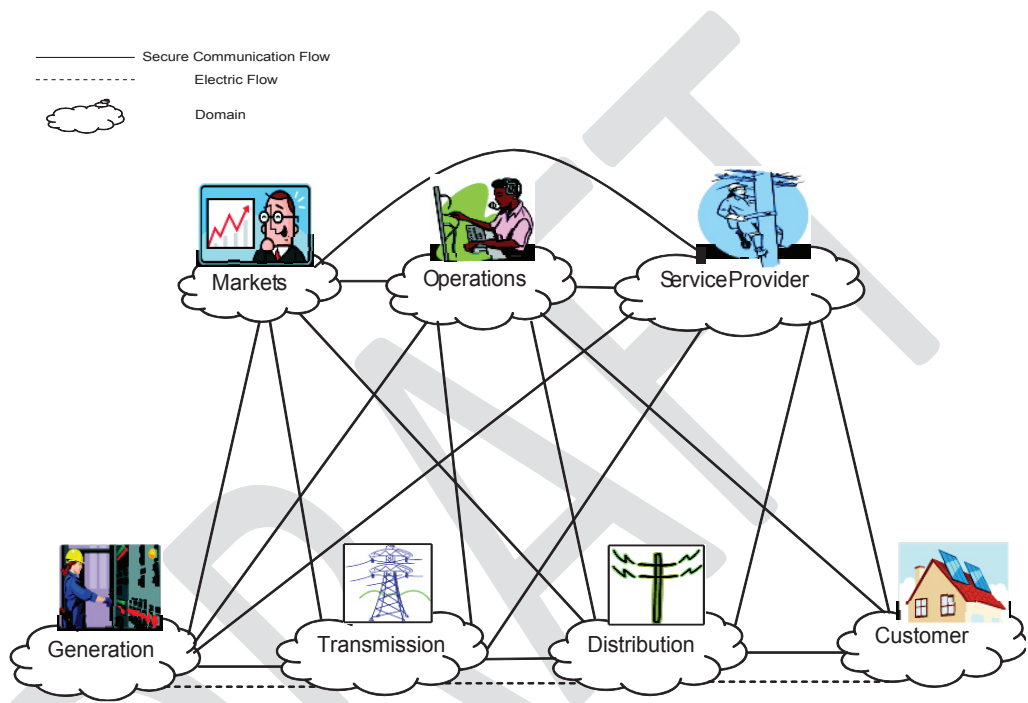

Figure: Typical smart grid architecture [11]

The backbone of a smart grid is the physical power system. In recent days, many new types of load are being introduced in the grid, including plug-in hybrid electric vehicles. Therefore, power demand is increasing rapidly and grid is becoming more complex due to the adoption of new technologies. Moreover, green technologies are increasingly used to make the grid more sustainable. These new ideas and technologies are making the grid more complex to analyse. In such a complex cyber-physical smart grid, the resources are coordinated by the control centre, which can be considered as the brain of the smart grid. These control centres are interconnected by a bidirectional cyber system including communication network, software and hardware. Consequently, the introduction of new technologies in the grid is making the cyber-physical system more vulnerable to the cyber threats that can degrade the performance of the physical system and even can cause a critical cascading failure of the power grid. The potential risks associated with the cyber-physical smart grid include [11]:

1. Increased complexity: The interconnection of new technologies in the grid is introducing greater complexities.

2. Risk of cascading failures: In a Smart Grid, the cyber system and the physical system are very tightly coupled. Therefore, failure due to a random attack or a targeted attack in either domain may affect the other domain and may lead to potential cascading failures.

3. Increase in potential adversaries: As the number of network nodes increase, the entry points for attackers on the network increase also increases introduces potential risks. This is considered as one of the main reasons why malicious code and related types of attacks and intrusions are also increasing for Smart Grid.

4. Data privacy issues: In a smart grid, widespread use of Intelligent Electronic Devices (IEDs) have increased the data gathering and two-way information flows extensively which have introduced the problems related to data confidentiality and intrusions of customer privacy.

# 4 Security Requirements of a Smart Grid

The security requirement of a Smart Grid is different from other critical infrastructures. The security objectives of the smart grid can be classified into three groups that are discussed below [11]:

## 4.1 Data Availability

This requirement of data security is very important and one of the primary objective to ensure the reliable operation of the smart grid. Generally, availability refers to the "timely and reliable access to the use of information" [11]. However, the time latency of the availability depends on the application as shown in Table. 1.

| Time Requirements | Data availability for the specific applications |
|---|---|
| <= 4 ms | Protective relaying |
| Sub seconds | Transmission wide-area situational awareness monitoring |
| Seconds | Substation and feeder SCADA data |
| Minutes | Monitoring noncritical equipment and some market pricing information |

| Hours | Meter reading and longer-term market pricing information |
| Days/Weeks/Months | Collecting long-term data such as power quality information |

Table 1: Time latency for different smart grid applications [11].

## 4.2 Data Integrity

Data integrity means source and quality of data is known and authenticated. The modification or destruction of original data leads to loss of data integration. Loss of data integration may occur due to the intrusion in the cyber domain by the attacker or disgruntled employees or by human error. Data integrity problem degrades the reliability of the system and with the increased system complexity, this problem is rising rapidly.

## 4.3 Data Confidentiality

This security requirement is important for privacy concern of the end users. Although it has a least impact on the smart grid reliability, importance is increasing with the deployment of advanced metering infrastructure and decentralized control within the grid.

# 5 Smart Grid Anomalies

## 5.1 Anomalies in the State Estimation Program

In order to fulfil the ever growing load demand, it is important to operate the power system at its maximum capacity. For a safe and reliable operation of the power system, operators need to monitor and control the system as it progresses through its various operating states. Different Intelligent Electronic Devices (IEDs) e.g., like the Remote Terminal Units (RTUs) are used to monitor the system states. However, these measurement data may be corrupted by an intruder or affected by noise or may be missing due to the sensor failure. Power system operators need to have confidence about the measurement data. For this reason, state estimation is widely used by the power system operators to calculate the system states. In addition, state estimation algorithms can detect any bad data and provide high accuracy estimation using limited measurement [12]. The concepts of state estimation for the reliable operation of a power system already exist, and have been successfully applied by power companies. In recent years, the importance of the security of state estimation algorithms is increasing with the growing complexity of the Smart Grid interconnections. In the recent researches, it has been shown that the smart grid state estimators are now highly vulnerable to the cyber attacks. From the literature of recent cyber security analysis of the state estimation program is basically related to 'False Data Injection Attack' [13-18] and 'Load Redistribution Attack' [19-20]. Both of these types of attacks are data integrity attack and are discussed next.

**5.1.1 False Data Injection Attack:** During power system operation, state estimation (SE) is important for Optimal Power Flow (OPF) operation, Contingency Analysis (CA), Automatic Generator Control (AGC) etc. A simple block diagram of a power system control centre is shown in Figure. 2 where it can be seen that SE plays a vital role for smooth operation of different Energy Management System (EMS) applications. Basically, OPF, CA, and AGC take the output SE data as an input to make the intelligent decision. For the processing purpose, SE receives data from the SCADA network. Due to the 'False Data Injection Attack', SCADA sends the SE wrong information and makes the smart grid vulnerable.

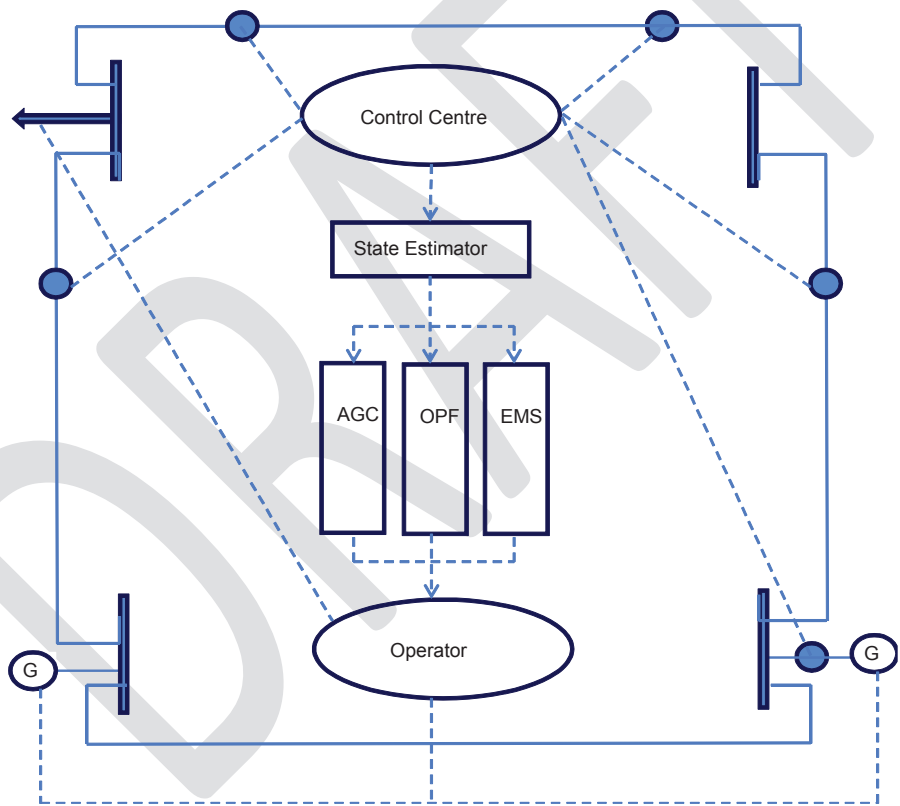

Figure 2: Energy Management System (EMS) working principles [16]

The states in a power system are the complex voltage magnitude and the angles of each bus. If the state vector is X, then

$$X = [\ \delta_1\ \delta_2\ \delta_3\ \ldots\ldots \delta_n\ \ V_1\ V_2 V_3 \ldots\ldots V_n]^T$$

Generally, the states of the system cannot be obtained directly; therefore, it is important to use the SE to infer the states from the measurement values. However, the measurement values may be noisy which increase the probability of error. As a result, SE can be traditionally formulated as the weighted least-square criterion below [14]:

$$\min J(x) = \sum_{i=1}^{m} w_i \left( z_i - h_i(x) \right)^2$$

where, h(x) is the measurement function which represents the measurement of z and w is the weight. Here m is the maximum number of the measurement. If there is no error in the measurement, then

$$z_i = h_i(x)$$

otherwise,

$$z_i = h_i(x) + e_i$$

where 'e' represents the error in the measurement. With the deployment of smart grid, SE is now vulnerable to cyber-attack. An intruder can attack on the measurement data of the SCADA system. Therefore, the control centre receives the following measured data due to the malicious data attack:

$$z_i = h_i(x) + e_i + \alpha$$

where, α is an attack vector. A significant number of researches are performed to prevent the false data intrusion which can be divided into three categories as below [13].

(i) Vulnerability Analysis of State Estimation
(ii) Consequence Analysis
(iii) Development of Countermeasures

The class of 'false data injection attack' in the electric power grid is first presented in [14] where author shows that an unobservable attack can be introduced based on a limited number of meters which can significantly degrade the performance of the results obtained from the SE [15]. Although bad data detection technique has been well established in the classical SE algorithm, malicious data attack is considered as the 'worst interacting bad data' injected by an intruder in [15]. A special type of stealth attack is discussed in [16] where the strategies of the defenders and the attackers are also investigated. In [17], the authors focus on the economic impact due to the false data injection attack on a real-time market operation of a power grid. In order to protect the grid from the false data injection, the strategic placement of Phasor Measurement Units (PMUs) is discussed in [18]. Efficient and accurate anomaly detection technique can be employed considering non-linear AC state estimation model to protect the grid from the intruders.

**5.1.2 Load Redistribution attacks in a power system**: Load Redistribution (LR) attack is a sub-class of false data injection attack. ED and OPF are heavily dependent on the output of the SE. Therefore, due to the LR attack, wrong estimation of the states may lead to an uneconomical solution and violate the stable operating conditions. Although a significant research has been done based on 'False Data Injection Attack', a few is performed considering LR attack model. LR attack model can be formulated as a bi-level programming problem [19]. This LR attack model has been formulated in [19] as below:

$$\sum_d \Delta D_d = 0$$
$$\Delta PL = -SF.KD.\Delta D$$
$$-\tau D_d \leq \Delta D_d \leq \tau D_d$$

The LR attack artificially increases or decreases demand at the load buses although total change of load remains zero as shown in the first equation. SF is the shifting factor matrix and KD is the bus-load incidence matrix. The attack magnitude of load $\Delta D_d$ is limited within an equality constraint as shown in the last equation. To solve the immediate LR attack problem, Karush-Kuhn-Tucker (KKT)-based method is proposed in [20]. Although KKT based method finds the global optimal solution, it is computationally very demanding. The efficiency of the proposed method is increased significantly in [19] using Benders decomposition. The concept of the most damaging LR attack is discussed from the attacker's perspective throughout the research in [19,20], significant contribution is yet to come considering power grid operator's or defender's perspective.

**5.2 Anomalies in the Power System Control Centres (PSCC)**

Generally, power system control centres receive information from the sensors using the SCADA network and it is the responsibility of the control centres to take intelligent decisions. The decision is then send to the actuators to perform actions on the field devices. A typical power system control loop is presented in Figure. 3. It is important to note that an adversary can easily exploit vulnerabilities along different steps of the control process of the power system [21]. The attack in the PSCC is related to the data integrity attack where information is corrupted, denial of service (DoS) attack, de-synchronization and timing based attacks [22].

Automatic Voltage Regulator (AVR) control, Governor Control (GC), and Automatic Generation Control (AGC) are essential in the generation side of a power system. Generally, the generation is controlled by local controller (e.g., AVR, GC) or wide area control (e.g., AGC) schemas [21].

**5.2.1 Attacks on the AGC:** In a power system, load is changing throughout the time. Therefore, AGC is used to balance the power output from different generation plants [23]. System frequency is monitored to take the decision of balancing the load demand and the generation of a system. In this secondary frequency control loop, both frequency and tie-line power are measured and send to other devices through wide area communication (e.g., IEC 61850) [21]. For control purpose, a point to point communication (e.g., DNP 3.0) is used [21]. As SCADA telemetry system is used for making the decision of AGC, security issues need to address to ensure a stable and reliable power grid operation.

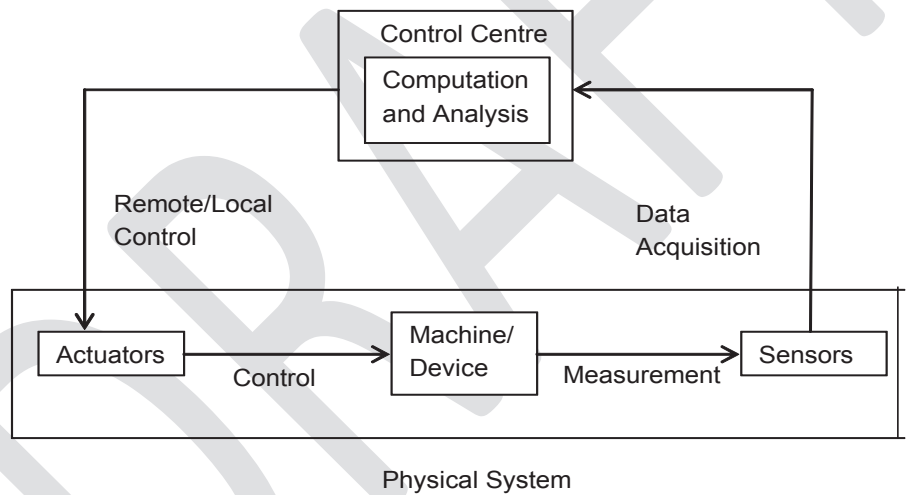

Figure 3: A typical power system control loop [10]

A reachability framework is developed in [24] to evaluate the impact of a two-area power system during a cyber attack. Based on the reachability framework, policy is developed that an attacker can follow to disrupt the power grid. Although the methodology of identification of AGC attack is proposed in [24], protection schemas need to be introduced for ensuring secured operation of AGC.

In [25], an approach to develop threat models for control system attacks is proposed. The proposed method in [25] is extended to define an attack model for a power system control centre by the authors in [26]. In that work, two type of attacks are considered which are 'min attack' and 'max attack'. The objective of these types of attacks is to manipulate the 'Area Control Error (ACE)' signal. Generally, the AEC is calculated from the difference of the net tie-line power flow to the deviation of frequency output [26]. To measure the actual power output and frequency sensors are used. During the attack, sensor measured values are manipulated which directly impacts on the system operating conditions.

**5.2.2 Attacks on GC:** Due to load incremental change of a generator, the electric output power exceeds the mechanical input power which leads to the speed deviation and frequency fall of the generator [27]. The reduction of generator speed is then detected by a sensor in a governor control system and necessary control actions are taken to run the generator in a steady-state condition. The GC is highly dependent on the local measurement. However, the modern generator governors use standard communication protocols to correspond information with the operation centre [21]. For example, the 800xA Governor of one the leading power and automation expert company ABB, makes the use of MODBUS, HART, PROFIBUS, PROFINET, DeviceNet, and IEC 61850 as a standard FieldBus and network communication protocol [28]. Therefore, an adversary can attack the GC system through any access point of the communication system. As GC plays a vital role for the stable operation of a generator, any cyber-attack can cause enormous disruption of the physical system.

## 5.3 Attacks on the FACTS device

Flexible Alternating Current Transmission Systems (FACTS) devices make the use of power electronics to stabilize and regulate power flow in a grid. Some applications of the FACTS devices are power flow control, load sharing, voltage regulation, transient stability enhancements, and power system oscillation mitigation [29]. FACTS devices help to utilize the network in a better way by increasing the capacity of the network with optimal power flow. However, optimal power flow of the network can not be achieved by a single FACTS devise and therefore multiple devices need to communicate and cooperate with each other during the operation. Therefore, communication link is important among the coordinated FACTS devices which also increase the vulnerabilities of cyber attacks [30]. To face the new challenges, two approaches are proposed in [30] which are agent-based management and improved visualization. Improved security policies and procedures need to address to handle the cyber attacks among the coordinated FACTS devices.

## 5.4 Malicious modification of network data stored in a database

Both the unauthorized access and malicious code is dangerous for stable power system operation. In [31], it has been reported that any of these types of attacks can disrupt the power system operation. To ensure the security of any power system control centre, requirements, policies and regulatory issues are pre-defined by the NERC, NIST and DOE [32] in the USA. However, alarmingly it has been shown [31] that any adversary can manage to get access to the network data stored in a database and manipulate the stored data leading to a compromise and failure of the Smart Grid.

Optimal Power Flow (OPF) is a widely adopted power system analysis tool used in the control centres for intelligent decision making. The operation of OPF is highly dependent on the network configuration data and the measured data obtained from the SCADA system. The interruption of power system operation due to the malicious modification of the network data stored in a database is discussed in [31]. A method based on Principle Component Analysis (PCA) is proposed to detect the anomalies related this type of attack. The method is successfully applied in IEEE benchmark test systems and has significant impact on false alarm reduction.

# 6 Protecting Smart Grid from cyber vulnerabilities

In recent years, the vulnerabilities of the smart grid has increased many times due to the wide adoption of communication network in different levels of operation and planning of a power grid. To protect the smart grid, it is important to protect the physical grid from three broad classes of cyber attacks [33] mentioned below.

## 6.1 Protection from component-wise cyber attack

To protect the smart grid at the device or component level, a security agent based framework has been proposed in [32, 34]. The security agents should be placed both in field devices (e.g., IEDs) and substation level (e.g., RTUs) as shown in Fig. 5. Some key functions of security agents described in [32] are:

1. Collecting network traffic patters and traffic data analysing

2. Maintaining data log and reporting

3. Run security patches and intrusion detection algorithms

4. Maintain end-to-end security

5. Alarm management

To protect the smart grid control system component, an anomaly detection technique has been proposed in [35-36]. In this research, a Rough Classification Algorithm has been used to detect the anomaly for improving the security of Power System Control Centres in the Electric Power System Critical Infrastructure. During the security analysis, authors have considered two operation modes (the normal and the abnormal operation mode) of the Smart Grid. The Rough Classification Algorithm is used for data reduction that enhances the performance of anomaly detection by introducing a compact set of knowledge-based rules. To protect the critical power system infrastructure a comprehensive framework has been developed in [37]. In this work, SCADA security has been investigated considering real-time monitoring, anomaly detection, impact analysis and implementing mitigation strategies as shown in Figure 5. In order to protect the relay from the false data attack, a probabilistic neural network based approach has been proposed in [38].

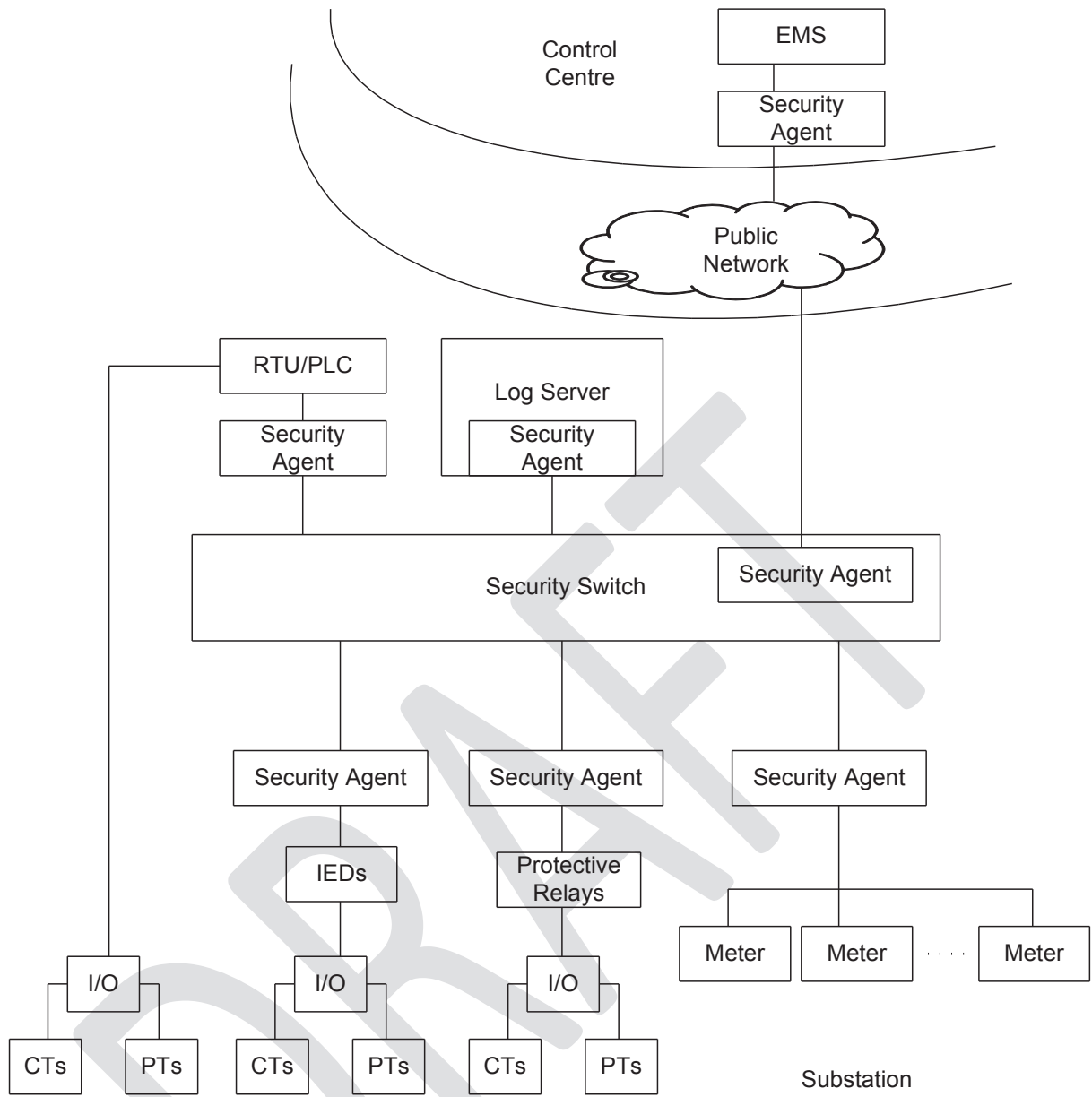

Figure 4: A security agent based framework for smart grid protection [33]

As smart meters and PMUs are vulnerable to cyber attacks, several research studies have been undertaken to learn about the security of the smart meters and PMUs. A strategic placement method of PMU components has been developed in [18]. The requirements and architectural directions of the IDS for smart meters are discussed in [39] where a specification based IDS is used. In the case of AMI security, specification based IDS shows its superiority over the signature based and anomaly based detection techniques due to: [39]

a. A greater accuracy of specification based IDS in AMI applications

b. the signature based IDS makes the use of black list approach which needs an attack data set. In terms of AMI, it is difficult to prepare an empirical attack data set.

c. the development of specification based IDS for AMI is cost effective.

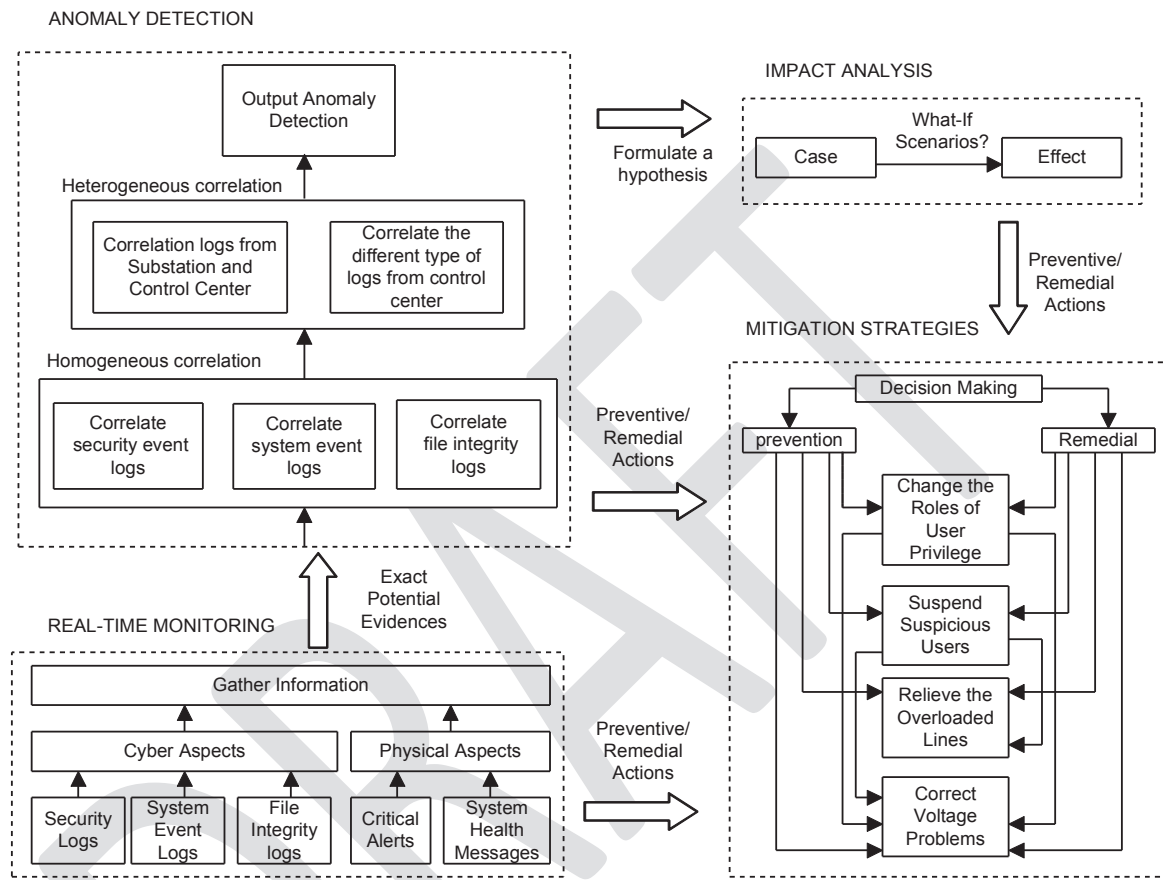

Figure 5: SCADA security framework based on real time monitoring and anomaly detection [37]

## 6.2 Protection from Protocol-wise Cyber Attack

SCADA is widely used in critical power system infrastructure. In case of multiple stakeholders, sometimes 'authentication', 'encryption' and 'firewalls' may not mitigate the security issues in a large SCADA network [40]. Moreover, focusing securing issues considering only SCADA network as a single entity will not solve the problem and therefore, it is important to ensure the cyber security of the individual devices in the network [40]. Different communication protocols are used among the SCADA devices for successful automation and operation of a smart grid. The evolution of the proprietary and industrial SCADA protocols started in early 1980s when Modbus, Modbus Plus, and proprietary and vendor specific protocols were first developed [41]. The Distributed Networking

Protocol (DNP) first appeared in 1990 by Westronic, Inc as an open protocol. The DNP3 protocol is based on the IEC 60870-5 protocol. Although DNP3 protocol is designed for reliable data communication, it is still vulnerable to cyber attack. Therefore, a rule based data set security for DNP3 devices is proposed in [40] to protect the smart grid from cyber terrorists. For simulating protocol attack, a SCADA simulation framework is developed in [42] where attack on a Modbus protocol is illustrated. Guidelines and the best practices for the development of smart grid protocols considering design principles are discussed in [43].

### 6.3 Protection from Topology-wise Cyber Attacks

Smart grid is also vulnerable to topology wise cyber attacks. For example, based on the knowledge of the power system topologies, an intruder may attack the bad data detection algorithms of the current state estimators [44]. Another topology based cyber attack is proposed in [45], where attack on the electric circuit breaker will cause the isolation of the generation units from the power grid. In [46], it has been shown that a cyber attack on confidentiality with proper topological knowledge can lead to an integrity and availability attack. Therefore, an information flow security-based model is proposed for mitigating these security issues. In [47], an optimum inter-link placement strategy against random attacks in the cyber-physical network is proposed which demonstrates that the strategy ensures a better security compared with all other possible strategies, including strategies using random allocation, unidirectional interlinks, in the case when topology of the cyber and physical network is unknown to each other.

## 7 Security Issues for Future Smart Grid

To make the grid smarter, significant initiatives are taken throughout the world. These measures will not only modernize the grid but also improve the overall system efficiency, stability and obviously reliability. But security issues must be maintained to ensure the uninterrupted power supply to the end users and to protect the national electricity grid from terrorist attacks. It is important to mention that a properly designed defence framework against cyber-attack should address all aspects related to the cyber-crime in a complex cyber-physical electricity grid infrastructure. That means, not only targeted cyber-attack should be considered but also, unintentional ICT related anomalies should be addressed, e.g., human operator errors, software errors, equipment failures and obviously natural disaster related problems.

In the process of making the power grid smarter, more automated control is being introduced in the grid. The risk of cyber attack will increase as the grid becomes more automated. Specially, control centres are the main target by the cyber terrorists. Energy utilities are applying advanced techniques and cyber security plans to avoid cyber-attacks. Advanced intrusion detection and prevention techniques can be implemented in different entry point of the complex grid. Security management systems are being implemented in different utilities. Energy providers are also adopting different risk-management strategies and defence approach against cyber-attack.

It is obvious that smart grid is providing lots of benefits including energy-efficient smart home, greener technology like solar and wind, cost-effective demand-side management, smart charging stations for electric cars and so on. In order to ensure these benefits, smart grid security measures must be maintained.

## 8 Conclusions:

In recent years, the numbers of cyber attacks are increasing rapidly. The intelligent cyber terrorists with detail and advanced power system knowledge may be able to create an integrity, availability or confidentiality attack on the network. Protection of smart grid from cyber attack is not only a concern of the engineers, researchers and the utility operators; it is also the responsibility of the government to ensure the security of this national critical infrastructure.

This chapter is written for the general readers so that they could be able to easily grasp some of the concepts in the area of Cyber Security for Smart Grid. At the beginning, a brief overview of Smart Grid and some recent Cyber Security incidents of this critical infrastructure are discussed. The key security requirements of a smart grid, which are 'availability', 'integrity', and 'confidentiality', are also discussed. Based on the existing research, an overview of Smart Grid anomalies is discussed thoroughly. The protection frameworks of smart grid against component-wise, protocol-wise and topology-wise cyber-attacks are also reviewed in this chapter.

Cyber security is very crucial for the reliable and secured operation of a critical smart grid infrastructure. At present, only Bad Data Detection (BDD) algorithms are used for data security in the state estimation. However, an adversary can attack the cyber-physical grid through any of the entry point of the cyber system and impact direly on the physical assets. For enhanced smart grid reliability and security, intrusion detection algorithms should be placed throughout system.

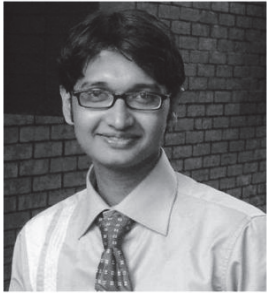

**Adnan Anwar** received the BSc degree in Electrical and Electronic Engineering from Islamic University of Technology (IUT), the organization of Islamic Conference. He has completed his Master by Research degree from the University of New South Wales, Canberra, Australia. He joined the University of Asia Pacific (UAP) as a lecturer in 2009. Currently, he is working with National ICT, Australia (NICTA) at the "Future Energy System" project. His research interest includes computational intelligence for the Smart Grid and its applications.

Adnan Anwar
Future Energy System
Optimization Research Group
National ICT, Australia
adnananwar@ieee.org

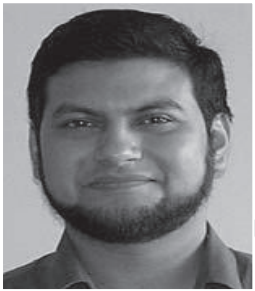

**Abdun Naser Mahmood** received the BSc degree in Applied Physics and Electronics and the MSc degree in Computer Science from the University of Dhaka, Bangladesh, in 1997 and 1999, respectively. He completed his PhD degree from the University of Melbourne in 2008. He joined the University of Dhaka as a lecturer in 2000, Assistant Professor in 2003, when he took a leave of absence for his PhD studies. Currently, he is working as a Lecturer at the University of New South Wales with the School of Engineering and Information Technology. His research interests include data mining techniques for network monitoring and algorithm design for anomaly detection and intrusion detection.

Abdun Naser Mahmood
School of Engineering and Information Technology
University of New South Wales
Canberra 2600, Australia
a.mahmood@adfa.edu.au